\documentclass[
    ,final            
  ]
  {aipproc}
\usepackage{psfrag,amsmath,graphicx}
\layoutstyle{6x9}

\begin{document}

\title{Aspects of unconventional density waves}

\author{ Kazumi Maki}{
  address={Department of Physics and Astronomy, University of Southern
California, Los Angeles CA 90089-0484, USA}
}
\author{Balázs D\'ora}{
  address={The Abdus Salam ICTP, Strada Costiera 11, I-34014, Trieste, Italy}
}
\author{ Attila Virosztek}{
  address={Department of Physics, Budapest University of Technology and
Economics, H-1521 Budapest, Hungary}
, altaddress={Research Institute for Solid State Physics and Optics, P.O.Box
49, H-1525 Budapest, Hungary}
}

\begin{abstract}
Recently many people discuss unconventional density waves (i.e. unconventional charge density 
waves (UCDW) and unconventional spin 
density waves (USDW)).
Unlike in conventional density waves, the quasiparticle spectrum in these systems is gapless. 
Also these systems remain metallic.
Indeed it appears that there are many candidates for UDW. The low temperature phase of 
$\alpha$-(BEDT-TTF)$_2$KHg(SCN)$_4$, the 
antiferromagnetic phase in URu$_2$Si$_2$, the CDW in transition metal dichalcogenite NbSe$_2$, the 
pseudogap phase in high $T_c$ 
cuprate superconductors, the glassy phase in organic superconductor $\kappa$-(BEDT-TTF)$_2$Cu[N(CN)$_2$]Br.
After a brief introduction on UCDW and USDW, we shall discuss some of the above systems, where 
we believe we have evidence for 
unconventional density waves.
\end{abstract}

\maketitle


\section{Introduction}

As is well known quasi-one dimensional electron systems have four canonical ground states: 
s-wave (spin-singlet)
superconductor, p-wave (spin-triplet) superconductor, charge density wave (CDW) and spin density
wave (SDW) \cite{solyom,gruner,lang,ishiguro}. All of these states have quasiparticle (QP) energy 
gap $\Delta$ and their QP density 
decreases 
exponentially at low temperatures ($T\ll \Delta$). Also the thermodynamics of these states is 
practically described by the BCS theory 
of s-wave superconductors\cite{BCS}.
Indeed except for p-wave superconductors these ground states have been found and their properties 
are actively persued even today.
As to p-wave superconductors it is most likely realized in quasi-one dimensional superconductor 
Bechgaard salts or (TMTSF)$_2$X with 
X=PF$_6$ and ClO$_4$\cite{lee}.
The thermal conductivity measurement show the presence of energy gap, and most recent NMR study 
indicates the triplet 
pairing\cite{lee}.

However, since 1979 a new class of superconductors have appeared on the scene: heavy fermion 
superconductors (1979), organic 
superconductors (1980),
high $T_c$ cuprate superconductors (1986), Sr$_2$RuO$_4$ (1994) and rare earth transition metal 
borocarbides (1994).
Now most of these new superconductors look like unconventional and/or nodal\cite{sigrist,szummad-wave}.
However only recently d$_{x^2-y^2}$-symmetry of both hole and electron doped superconductors have 
been established\cite{revmod}. Also the 
superconductivity in Sr$_2$RuO$_4$ is f-wave\cite{izawa,optical,raman} and the one in YNi$_2$B$_2$C
 and LuNi$_2$B$_2$C is 
s+g-wave\cite{maki1,izawa2}. Therefore unconventional 
superconductivity has taken center stage in the 21st century physics.

Parallel to these developments many people consider unconventional and/or nodal density waves 
(UDW)\cite{Ners1,benfatto,nagycikk,nayak}. We believe now that the 
low-temperature 
phase (LTP) of $\alpha$-(BEDT-TTF)$_2$KHg(SCN)$_4$\cite{ltp,rapid,physicaB,alfa}, the 
antiferromagnetic phase in 
URu$_2$Si$_2$\cite{IO,roma}, the CDW in 2H-NbSe$_2$\cite{castroneto}, the
pseudogap 
phase in high $T_c$ cuprates\cite{benfatto,nayak,app} and the glassy phase in 
$\kappa$-(ET)$_2$ salts\cite{pinteric} belong to UDW.

\section{Physical properties of UCDW and USDW}

First of all the thermodynamic properties of UDW are very well described in terms of mean field 
theory like the BCS one.
In fact the thermodynamics of most of UDW is described in terms of the BCS theory for d-wave 
superconductors\cite{szummad-wave,d-wave}. 
Qualitatively the 
thermodynamics of d-wave superconductors is not much different from the one for s-wave superconductors.
In particular a clear jump in the specific heat at $T=T_c$ (transition temperature)  is observable 
in both cases.
On the other hand at low temperatures, unlike in conventional DW, there are nodal excitations, 
giving rise to the power law specific 
heat like $C\sim T^2$.
Also due to the nodal excitations UCDW and USDW are metallic down to $T=0$K.
Further unlike conventional density wave there is no clear x-ray or spin signal indicating 
the phase transition, since $\langle \Delta({\bf k})\rangle=0$. 
Here $\langle \dots \rangle$ means average over the Fermi surface. 
For this reason UDW is an important 
candidate for states with hidden order parameter.

For the existence of UDW we need higher dimensionality and competing interactions. 
Therefore we can see here clearly the paradigm shift 
from quasi-one dimensional systems to quasi-two dimensional and three dimensional systems.
Also in order to study UDW experimentally we need more subtle and delicate technique. In this context
 the angular dependent 
magnetoresistance provides a unique window to study UDW.

\section{Angular dependent magnetoresistance in $\alpha$-(BEDT-TTF)$_2$KHg(SCN)$_4$}

\begin{figure}[h!]
\psfrag{a}[t][b][1][0]{$a$}
\psfrag{b}[][][1][0]{$c$}
{\includegraphics[width=5cm,height=5cm]{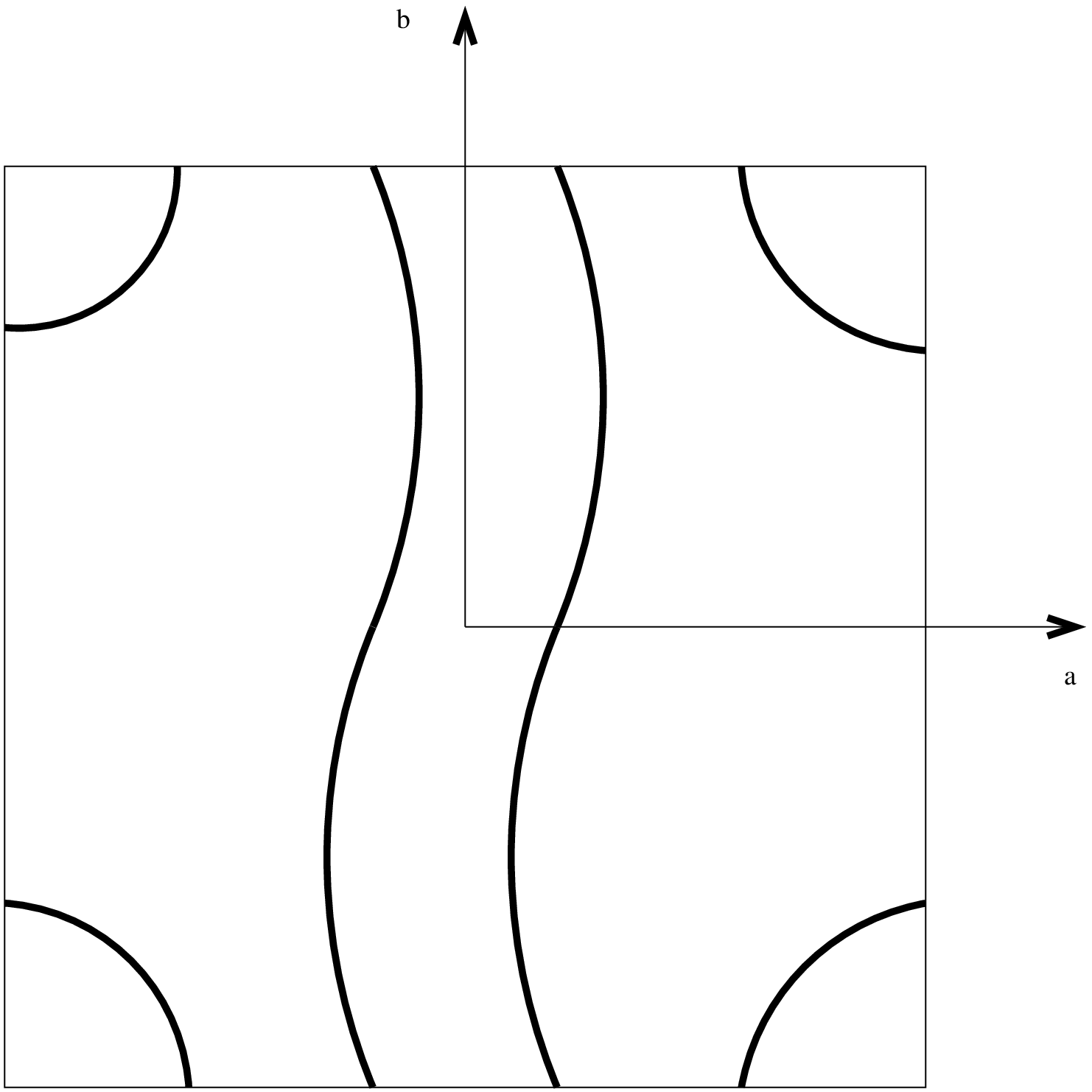}}
\label{fermisurf}
\hspace*{10mm}
\psfrag{B}[bl][tr][1][0]{$\bf B$}
\psfrag{c}[b][t][1][0]{$c$}
\psfrag{a}[b][t][1][0]{$a$}
\psfrag{b}[r][l][1][0]{$b$}
\psfrag{pp}[][][1][0]{$\phi$}
\psfrag{p}[][][1][0]{$\theta$}
\includegraphics[width=4cm,height=4cm]{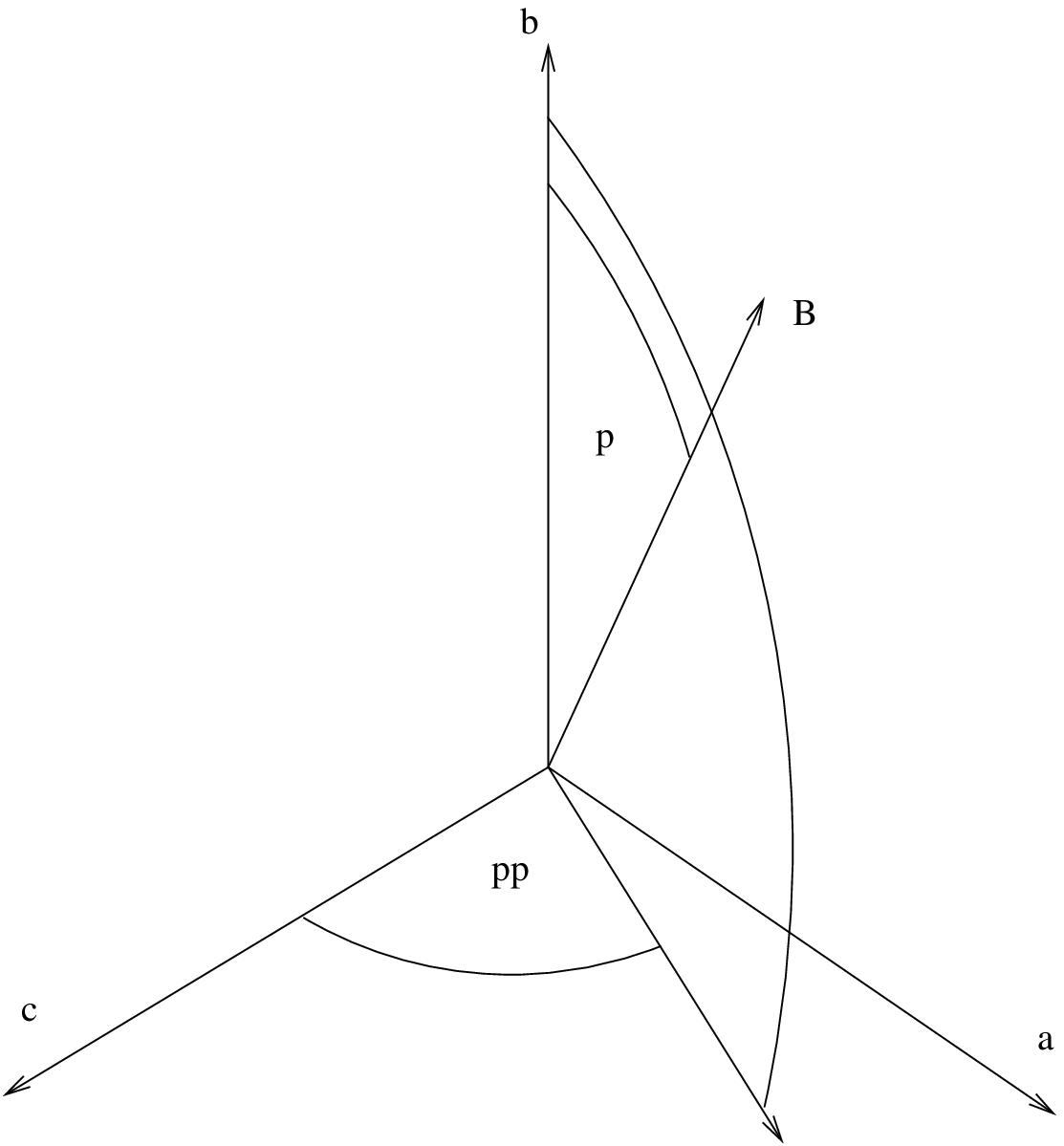}
\caption{The Fermi surface of $\alpha$-(BEDT-TTF)$_2$KHg(SCN)$_4$ is shown in the left panel. In the right one the
geometrical
configuration of the magnetic field with respect to
 the conducting plane is plotted.}
\end{figure}

\begin{figure}[h!]
\psfrag{x}[t][b][1][0]{$\theta$ ($^\circ$)}
\psfrag{y}[b][t][1][0]{$R_\parallel(15T,\theta)$ (Ohm)}
\includegraphics[width=68mm,height=68mm]{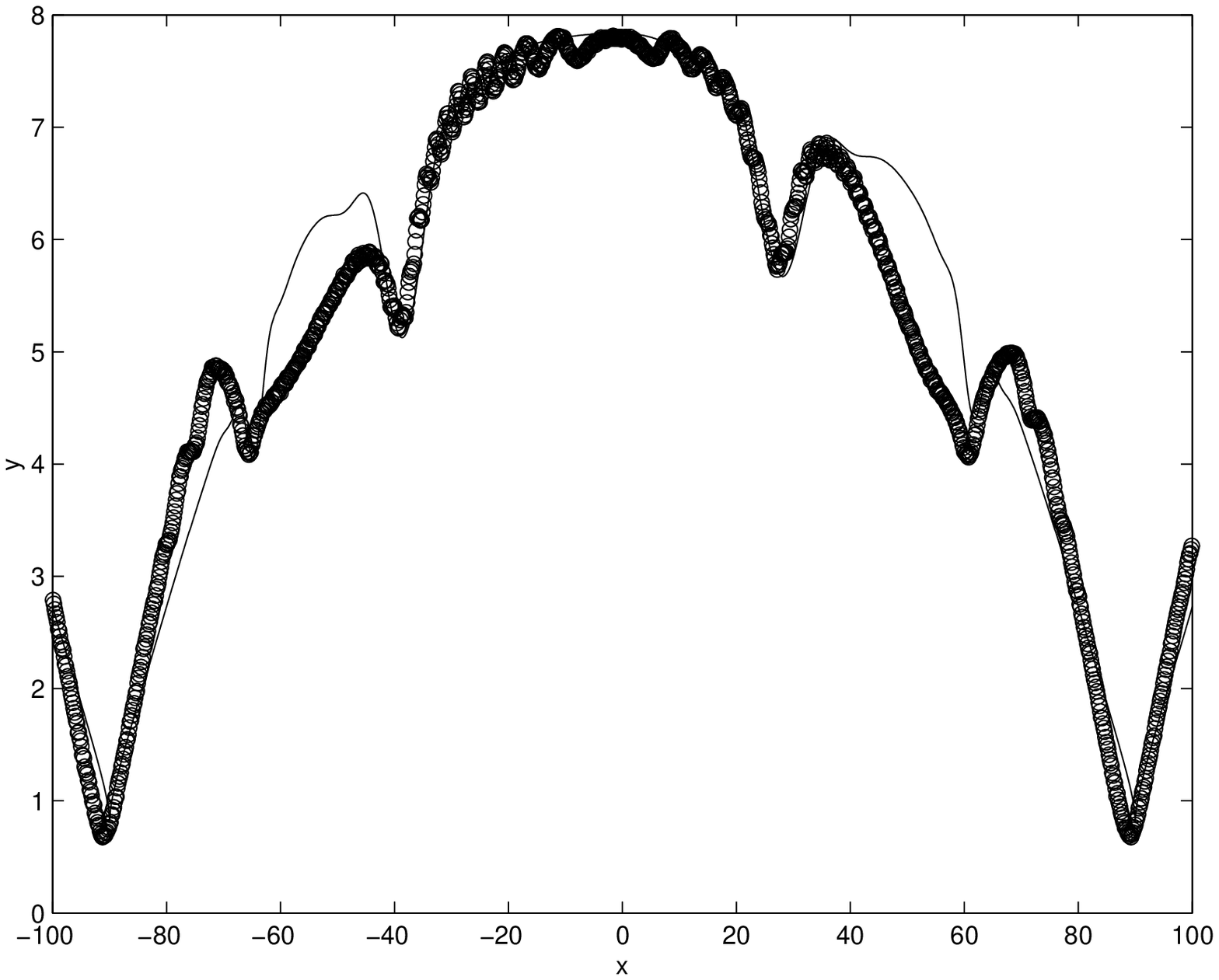}
\hspace*{2mm}
\psfrag{x}[t][b][1][0]{$\theta$ ($^\circ$)}
\psfrag{y}[b][t][1][0]{$R_\perp(15T,\theta)$ (Ohm)}
\includegraphics[width=68mm,height=68mm]{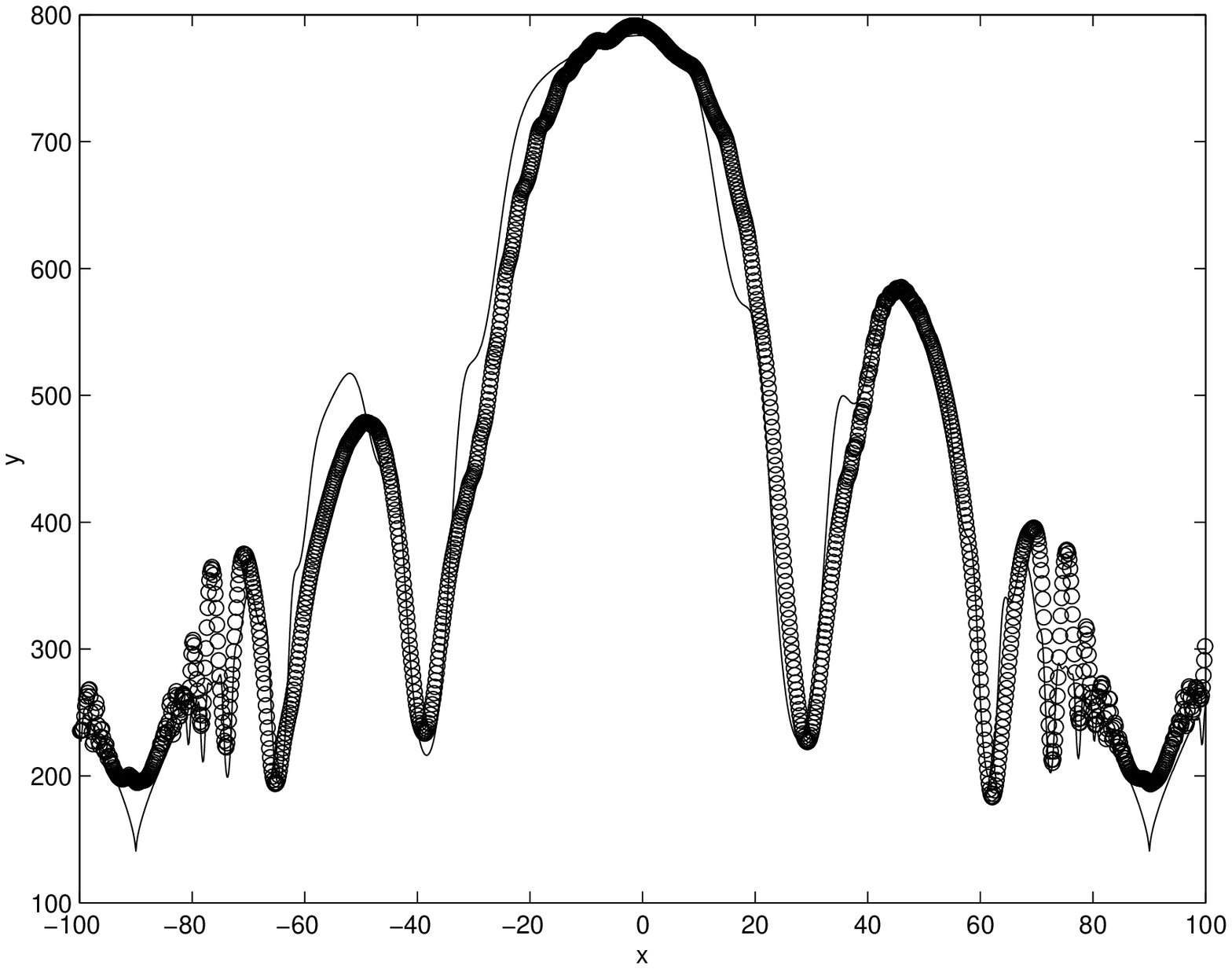}

\caption{The angular dependent magnetoresistance is shown for current parallel (left panel) and perpendicular (right panel)
to the $a-c$ plane at $T=1.4$K,
$B=15$T, $\phi=45^\circ$. The open circles belong to the experimental
data, the solid
line is our fit based on Eq. (\ref{fit}).}
\label{rperp}
\end{figure}

The LTP in $\alpha$-(BEDT-TTF)$_2$MHg(SCN)$_4$ with M=K, Tl, Rb is still controversial. This 
compound is quasi-two dimensional system 
with 1D like and 2D like Fermi surfaces as shown in Fig. \ref{fermisurf}\cite{singl}. From the 
magnetic phase diagram in a magnetic 
field ${\bf 
H}\parallel b^*$, it is believed that the LTP is not SDW but a kind of CDW\cite{karts1}. We have 
proposed recently that UCDW can 
account for a 
number 
of features in LTP of $\alpha$-(BEDT-TTF)$_2$KHg(SCN)$_4$ including the threshold electric 
field\cite{ltp,rapid,tesla,imperfect}. More 
recently we have discovered that 
the angular dependent magnetoresistance (ADMR) observed in LTP can be interpreted in terms of Landau 
quantization of the quasiparticle 
spectrum in UCDW\cite{Ners1,alfa,prl}.
First let us assume that the QP spectrum in UCDW is given by\cite{alfa,prl}
\begin{equation}
E({\bf k})=\sqrt{\xi^2+\Delta^2({\bf k})}-\varepsilon_0\cos(2\bf b^\prime k),\label{elso}
\end{equation}
where $\xi\approx v_a(k_a-k_F)$, $v_a$ is the Fermi velocity, $\Delta({\bf k})=\Delta\cos(ck_z)$, $\bf b^\prime$ is the vector lying
outside of the $a-c$ plane and
$\varepsilon_0$ is the parameter describing the imperfect nesting\cite{yamaji1,yamaji2,huang3,physicaB}.
In fitting the experimental data we discovered that 1. Eq. (\ref{elso}) gives only one single dip
in ADMR, 2. therefore the imperfect nesting term has to be generalized as
\begin{equation}
\varepsilon_0\cos(2{\bf b^\prime k})\longrightarrow \sum_n\varepsilon_n\cos(2{\bf
b}^\prime_n\bf k),
\label{generalization}
\end{equation}
where ${\bf b}^\prime_n=b^\prime[\hat{\bf
r}_b+\tan(\theta_n)(\hat{\bf r}_a\cos\phi_0+\hat{\bf r}_c\sin\phi_0)]$, $\varepsilon_n=\varepsilon_0 2^{-|n|}$,
$\tan(\theta_n)=\tan(\theta_0)+n d_0$. Indeed ADMR has a broad peak at ${\bf H}\parallel b^*$ (or $\theta=0$) and exhibits a number of 
dips at $\theta=\theta_n$ (see Fig. \ref{rperp}!)
\begin{equation}
\tan(\theta_n)\cos(\phi-\phi_0)=\tan(\theta_0)+nd_0,
\end{equation}
where $\tan\theta_0\simeq 0.5$, $d_0\simeq 1.25$, $\phi_0\simeq 27^\circ$
and $n=0$, $\pm1$, $\pm2$\dots\cite{fermi,kovalev}.
Now in the presence of magnetic field $\bf H$ with the orientation described by $\theta$ and $\phi$ (see Fig. \ref{fermisurf}), the 
QP spectrum 
changes to 
\begin{equation}
E_n=\pm\sqrt{2nv_a\Delta c e|B\cos\theta|},
\label{ketto}
\end{equation}
where $n=0$, $1$, $2$\dots.
This  is readily obtained following Ref. \cite{Ners1}.
The contribution from the imperfect nesting term is considered as a perturbation and the lowest order corrections to the
energy spectrum are given by:
\begin{gather}
E_0^1=E_1^1=-\sum_m\varepsilon_m\exp(-y_m),\\
E_1^2=-\sum_m\varepsilon_m(1-2y_m)\exp(-y_m),
\end{gather}
where $y_m=v_a {b^\prime}^2e |B\cos(\theta)|[\tan(\theta)\cos(\phi-\phi_o)-\tan(\theta_m)]^2/\Delta c$.
The $n=1$ level was twofold degenerate, but
the imperfect nesting term splits the degeneracy by $E_1^1$ and $E_1^2$. Also the imperfect nesting term breaks the particle-hole
symmetry.
When $\beta E_1\gg 1$ ($\beta=(k_BT)^{-1}$), the quasiparticle transport in the quasi-one dimensional Fermi surface
is dominated by the quasiparticles at $n=0$ and $n=1$ Landau levels. Considering that there are 2 conducting channels
and only the quasi-one dimensional one is affected by the appearance of UCDW, the ADMR is written as
\begin{eqnarray}
R(B,\theta,\phi)^{-1}=2\sigma_1\left(\dfrac{\exp(-\beta
E_1)+\cosh(\beta E_1^1)}{\cosh(\beta
E_1)+\cosh(\beta E_1^1)}+\dfrac{\exp(-\beta
E_1)+\cosh(\beta E_1^2)}{\cosh(\beta
E_1)+\cosh(\beta E_1^2)}\right)+\sigma_2
\label{fit}
\end{eqnarray}
Here $\sigma_1$ and $\sigma_2$ are the conductivities of the $n=1$ Landau level and quasi-two dimensional channels, in which the
contribution of the $n=0$ Landau level was melted,
respectively. The same expressions were found for $\Delta({\bf k})=\Delta\sin(ck_z)$.

\begin{figure}[h]
\psfrag{x}[t][b][1.2][0]{$T$(K)}
\psfrag{y}[b][t][1.2][0]{$R$(Ohm)}
\psfrag{m3}[l][r][1][0]{$B=15$T}
\includegraphics[width=68mm,height=68mm]{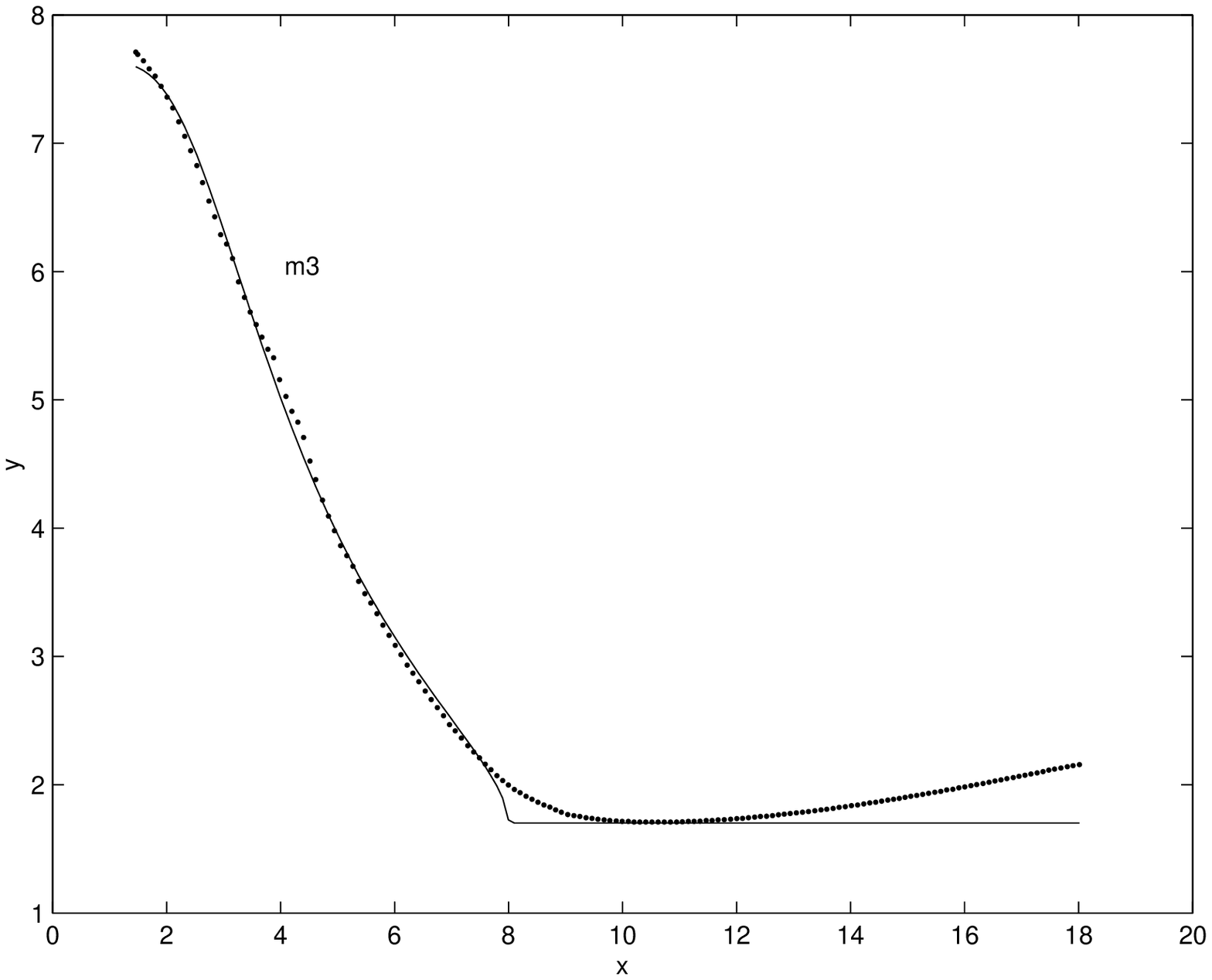}
\hspace*{2mm}
\psfrag{x}[t][b][1.2][0]{$B$(T)}
\psfrag{y}[b][t][1.2][0]{$R$(Ohm)}
\psfrag{m1}[t][b][1][0]{$T=1.4$K}
\psfrag{m2}[][][1][0]{$T=4.14$K}
\includegraphics[width=68mm,height=68mm]{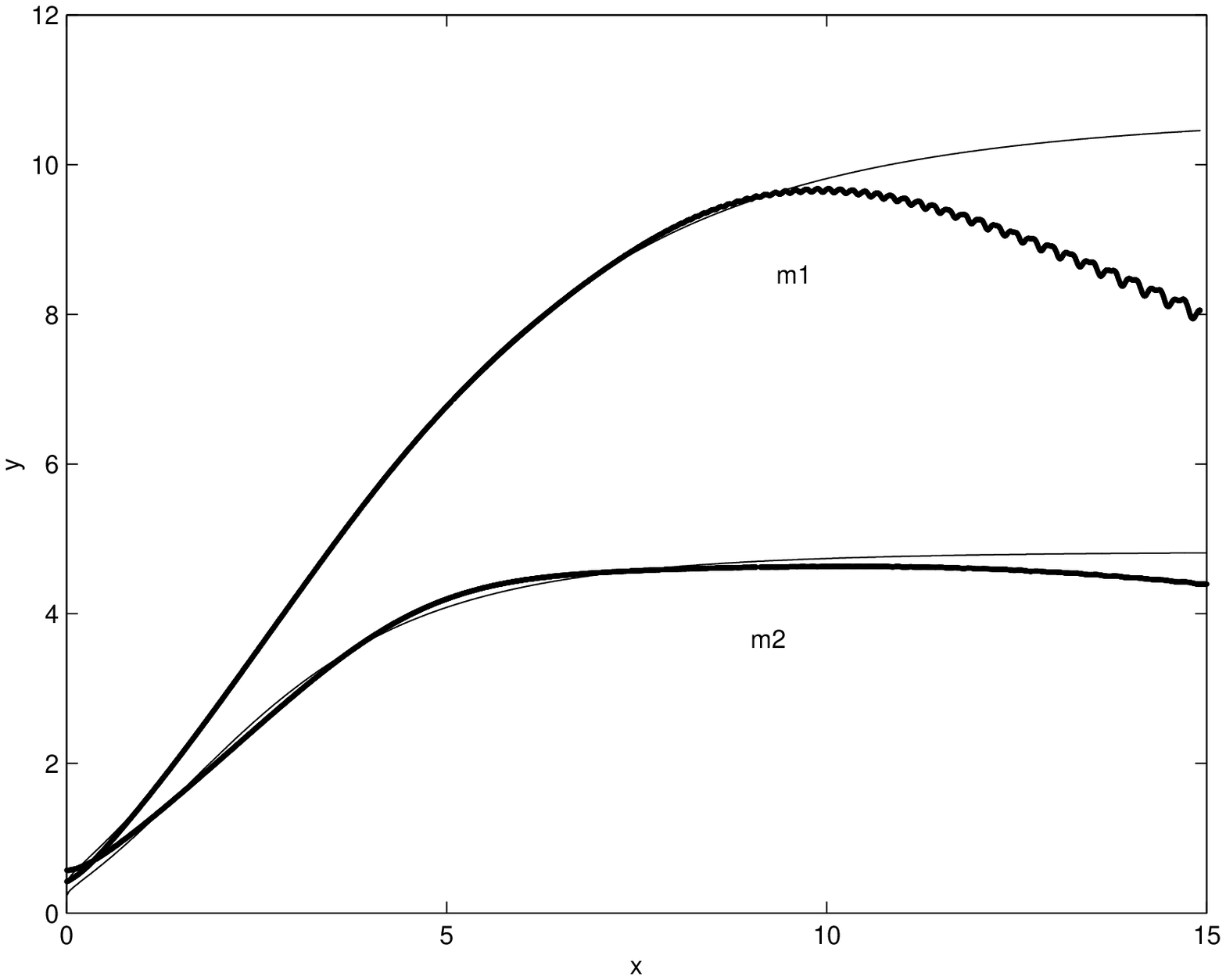}
\caption{The temperature dependent magnetoresistance is shown at $B=15$T in the left panel. The dots are the experimental data, the
solid line is our fit. In the right panel, the magnetoresistance is plotted for $T=1.4$K and $4.14$K as a function of magnetic field. 
The thick solid line is the
experimental data, the thin one denotes our fit based on Eq. (\ref{fit}).}\label{fig:koord}
\end{figure}

Eq. (\ref{fit}) is compared to the ADMR data taken from a single crystal of $\alpha$-(BEDT-TTF)$_2$KHg(SCN)$_4$ for the temperature 
interval 1.4-20~K under magnetic field up to 15~T\cite{prl}. The ADMR data are consistent with the previous 
reports\cite{kovalev,caulfield2,hanasaki}. 
In Fig. \ref{fig:koord} we compare the $B$ dependence of the magnetoresistance at $T=1.4$~K and
$T=4.14$~K and the $T$ dependence of the magnetoresistance for $B=15$~T, for $\theta=0^\circ$. In fitting the temperature dependence
of the resistivity, we assumed $\Delta(T)/\Delta(0)=\sqrt{1-(T/T_c)^3}$, which was found to be very close
to the exact solution of $\Delta(T)$\cite{nagycikk}. The influence of imperfect nesting terms in these cases is negligible, since they
contribute only close to $\theta=\theta_n$.

\begin{figure}[h!]
\psfrag{x}[t][b][1][0]{$\theta$ ($^\circ$)}
\psfrag{y}[b][t][1][0]{$R_\perp(15T,\theta)$ (Ohm)}
\includegraphics[width=68mm,height=68mm]{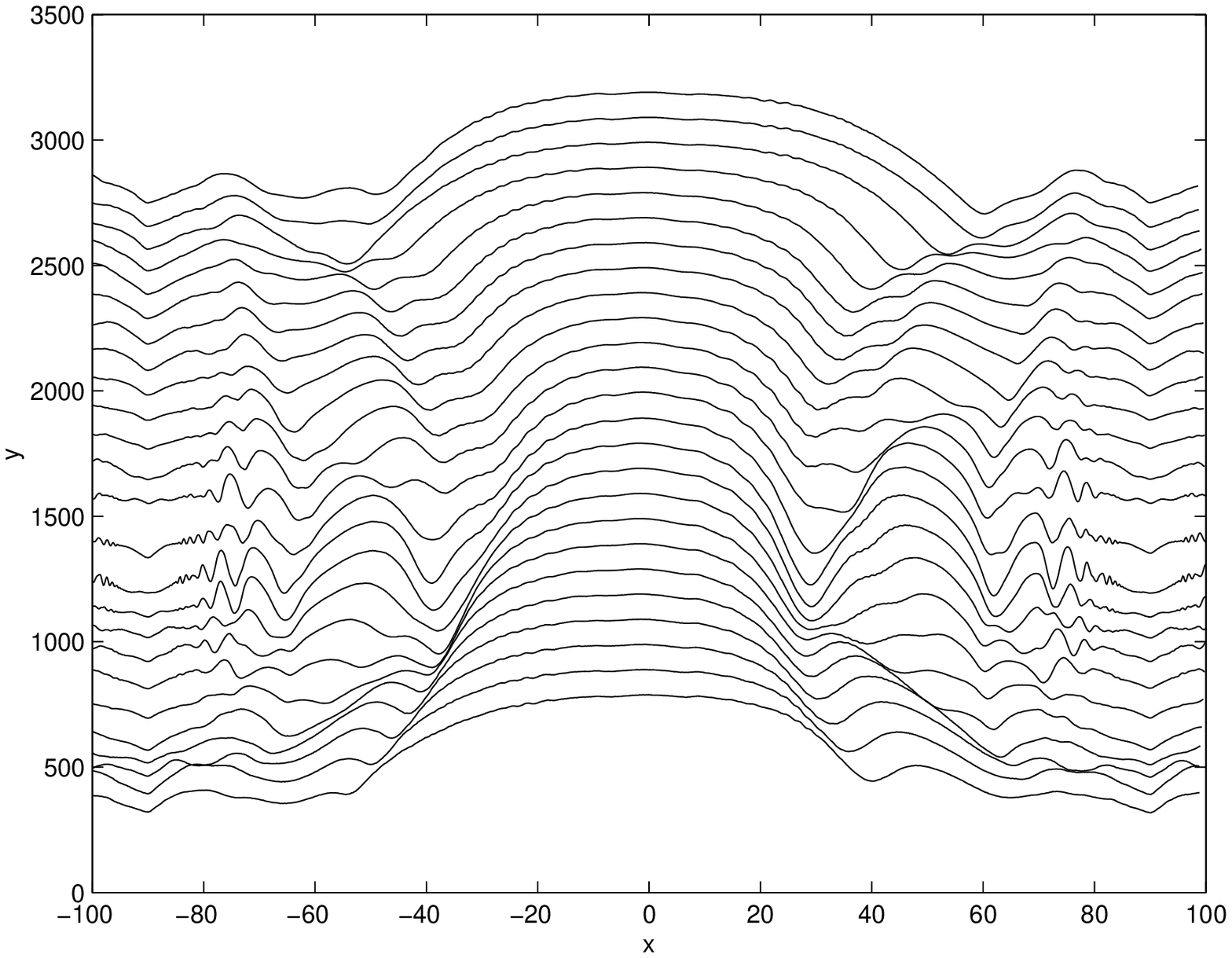}
\hspace*{2mm}
\psfrag{x}[t][b][1][0]{$\theta$ ($^\circ$)}
\psfrag{y}[b][t][1][0]{$R_\perp(15T,\theta)$ (Ohm)}
\includegraphics[width=68mm,height=68mm]{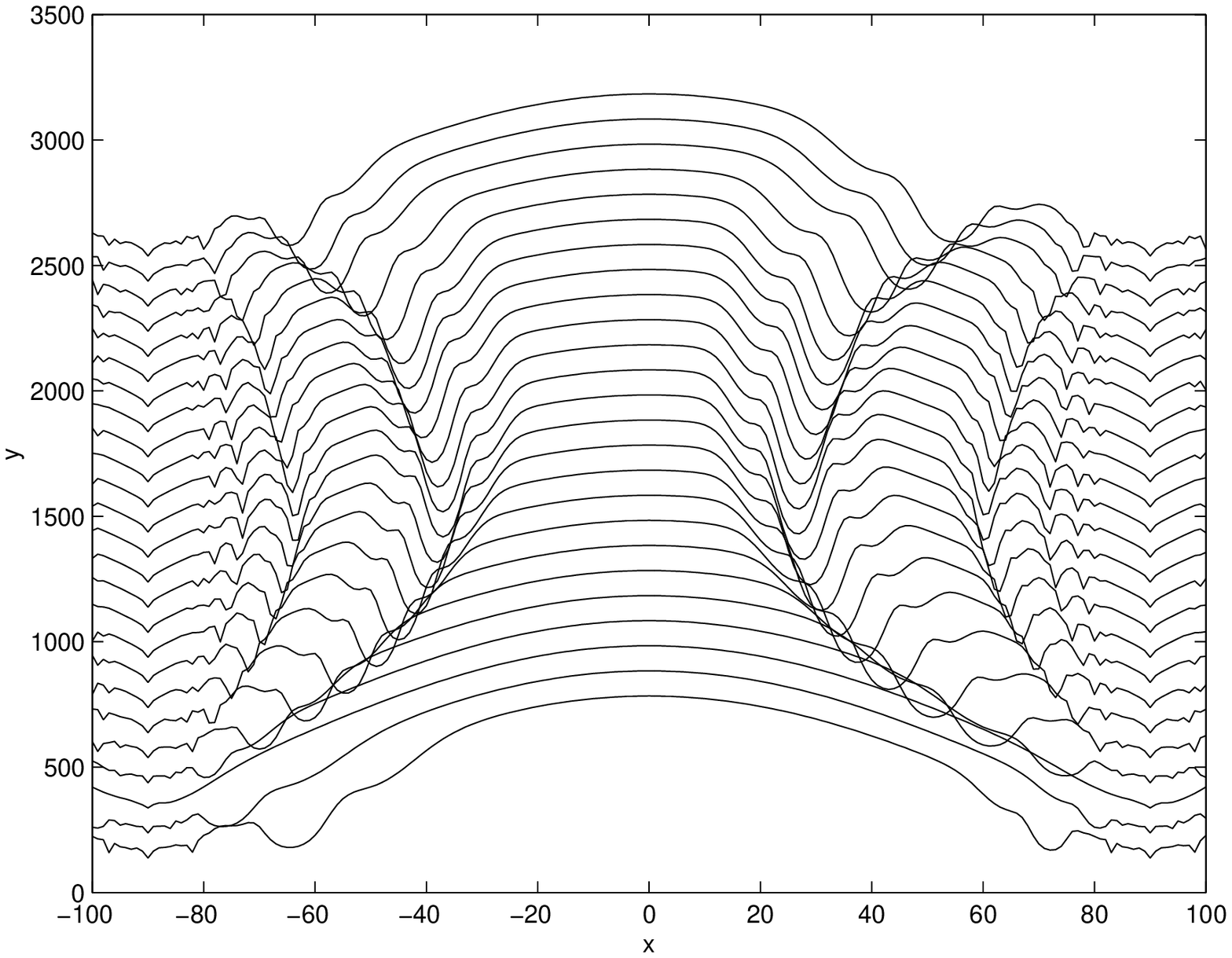}
\caption{ADMR is shown for current perpendicular to the $a-c$ plane at
$T=1.4$K and $B=15T$ for $\phi=-77^\circ$, $-70^\circ$,
$-62.5^\circ$, $-55^\circ$, $-47^\circ$, $-39^\circ$, $-30.5^\circ$, $-22^\circ$, $-14^\circ$, $-6^\circ$, $2^\circ$, $10^\circ$,
$23^\circ$, $33^\circ$, $41^\circ$, $48.5^\circ$, $56^\circ$, $61^\circ$, $64^\circ$,
$67^\circ$, $73^\circ$, $80^\circ$, $88.5^\circ$, $92^\circ$ and $96^\circ$ from bottom to top. The left (right) panel shows
experimental
(theoretical) curves, which are shifted from their original position along the vertical axis by $n\times100$Ohm, $n=0$ for 
$\phi=-77^\circ$, $n=1$ for $\phi=-70^\circ$, \dots.}\label{sokphit}
\end{figure}

Clearly the fitting becomes better as $T$ decreases and/or $B$ increases. Also for
$T=1.4$~K Shubnikov-de Haas oscillation becomes visible around $B=10$~T, then the fitting starts breaking away.
Clearly in this high field region the quantization of Fermi surface itself starts interfering
with the quantization described above. 
Also the deviation of the theoretical curve from the experimental one above $T_c$ in Fig. \ref{fig:koord} is due to the fact 
that the higher Landau levels contribute in this high temperature regime.
From these fittings we can deduce $\sigma_2/\sigma_1\sim 0.1$ and $0.3$, and by assuming the
mean field value of $\Delta$ ($17$~K), we get $v_a\sim 6\times 10^6$~cm/s.
In Fig. \ref{rperp}  we show the experimental data of ADMR as a function of $\theta$ for current parallel and
perpendicular to the conducting plane for $T=1.4$~K, $B=15$~T and $\phi=45^\circ$.
As is readily seen the fittings are excellent. From this we deduce $\sigma_2/\sigma_1\sim 0.1$,
 $b^\prime\sim 30$~\AA, $\varepsilon_0\sim 3$~K. This $b^\prime$ is comparable to the lattice constant $b=20.56$~\AA.
Finally we show in Fig. \ref{sokphit} $R$ versus $\theta$ for different $\phi$ and compare with the experimental data side by
side. Perhaps there
are still differences in some details but the overall agreement is very striking. 
The present model can describe a similar figure found in Ref. \cite{hanasaki} rather well.

In summary the Landau quantization of the QP spectrum of UDW as proposed by Nersesyan et al.\cite{Ners1} can account for the striking 
ADMR found in LTP
 of $\alpha$-(BEDT-TTF)$_2$KHg(SCN)$_4$. Very similar ADMR have been seen also in M=Rb and Tl compounds. Therefore we conclude that
 LTP in
$\alpha$-(BEDT-TTF)$_2$MHg(SCN)$_4$ salts should be UCDW. Also we believe that ADMR provides clear signature for
the presence of UCDW and USDW. Therefore this technique can be exploited for other possible candidates of UDW.

\section{Pseudogap phase in high $T_c$ cuprates}

\begin{figure}[h!]
\hspace*{8mm}

\psfrag{x}[t][b][1][0]{$x$}
\psfrag{0.05}[t][b][1][0]{$0.05$}
\psfrag{0.15}[t][b][1][0]{$0.15$}
\psfrag{0.25}[t][b][1][0]{$0.25$}
\psfrag{oo}[][][1][0]{$0$}
\psfrag{100}[][][1][0]{$100$}
\psfrag{500}[][][1][0]{$500$}
\psfrag{T}[r][l][1][0]{$T(K)$}
\psfrag{af}[][][1][0]{AF}
\psfrag{pseudo}[][r][1][0]{ pseudogap phase}
\psfrag{d-wave}[][r][1][0]{ d-wave SC}
\psfrag{t1}[][][1][0]{$T^*$}
\psfrag{t2}[][][1][0]{$T_c$}
{\includegraphics[width=7cm,height=68mm]{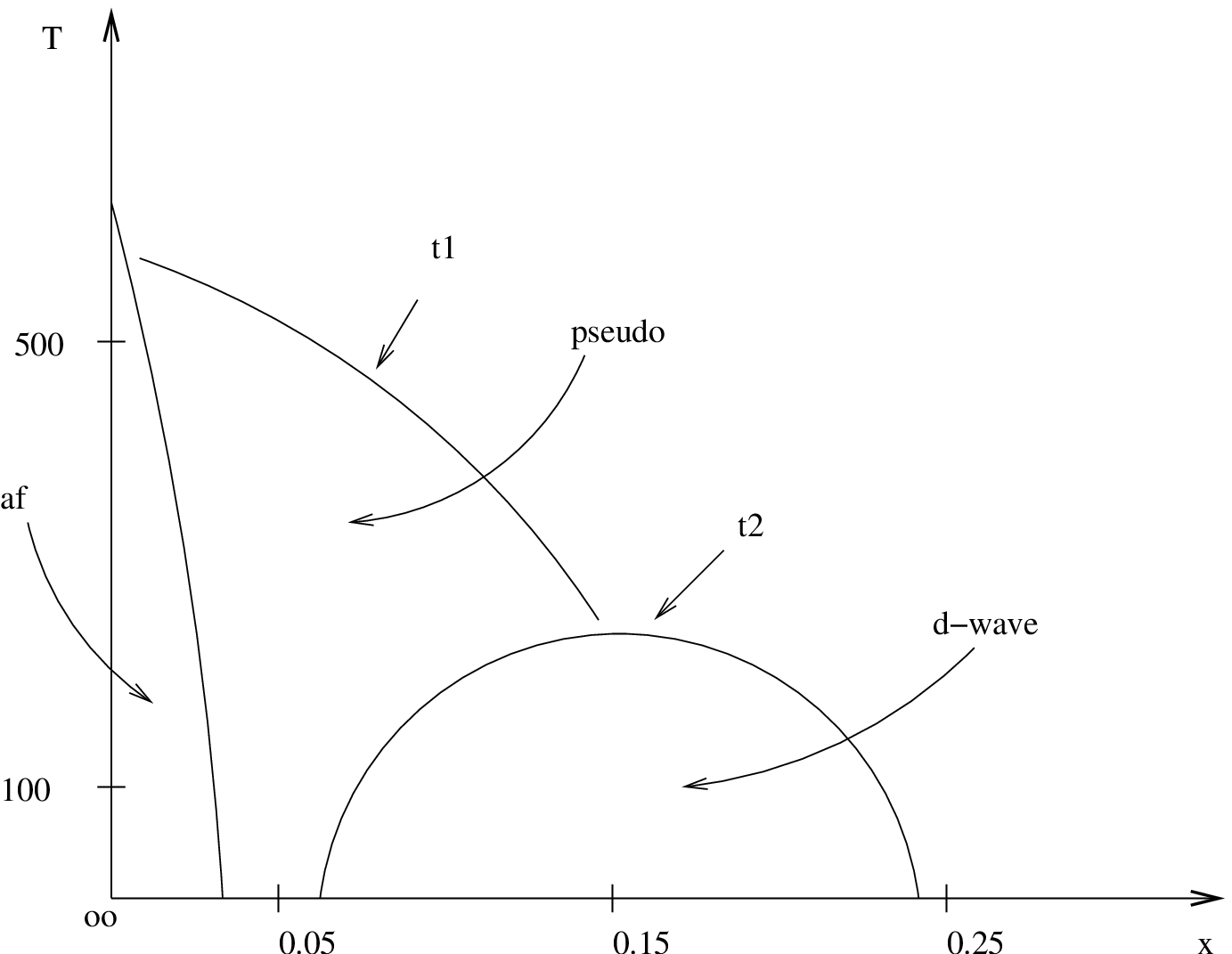}}
\hspace*{-3mm}
\vspace*{5mm}
\psfrag{x}[t][b][1][0]{$\theta$ ($^\circ$)}
\psfrag{y}[b][t][1][0]{$\rho(H,\theta,\phi)/\rho(H,0,0)$)}
\includegraphics[width=8cm,height=78mm]{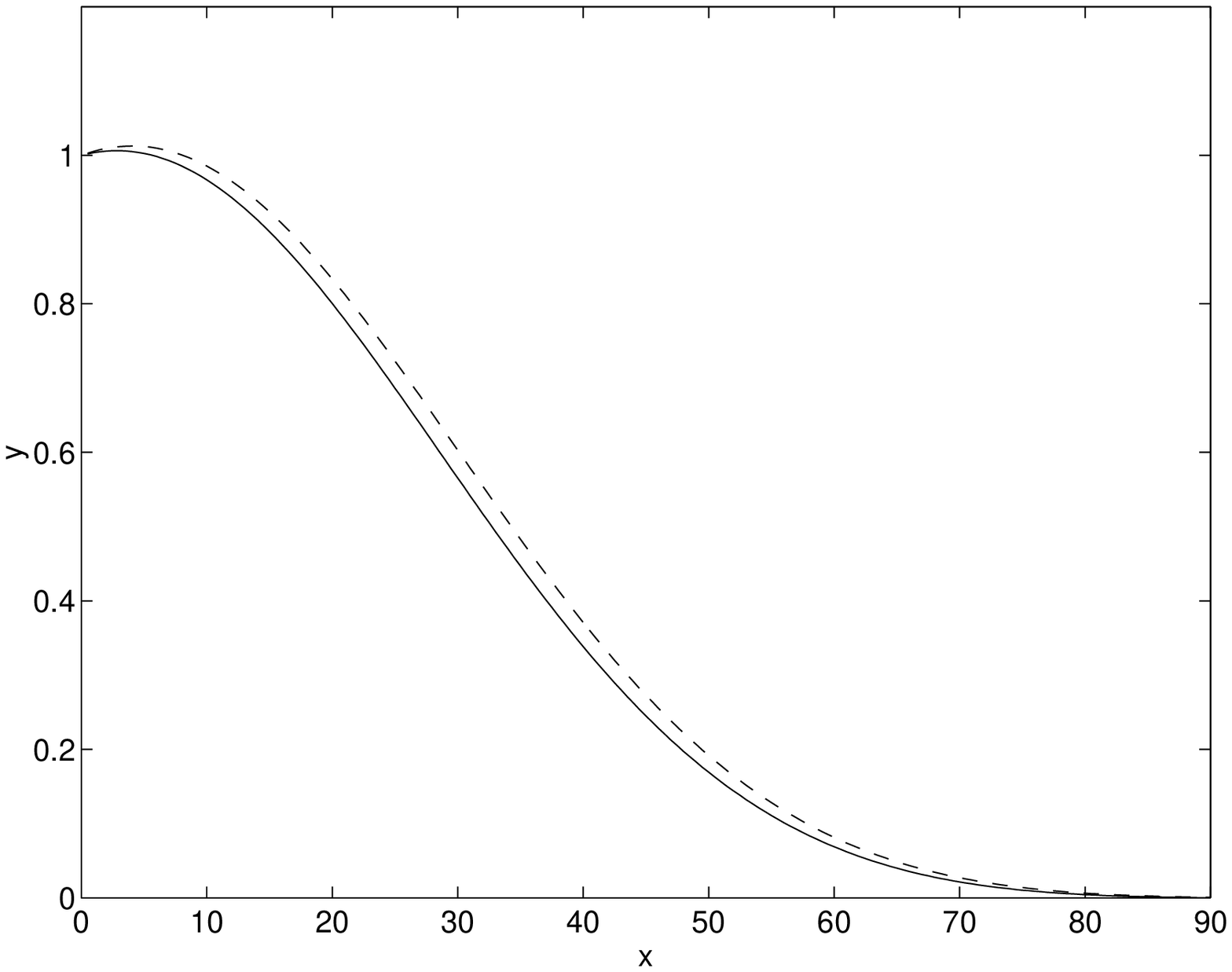}

\caption{Left panel: The schematic phase diagram of high $T_c$ cuprates. Right panel: The angular dependent
magnetoresistance is shown as a function of
$\theta$ for current parallel
to the $a$-axis for $\phi=0^\circ$ (dashed line) and $\phi=45^\circ$ (solid line)}
\label{phase}
\end{figure}

We believe that the most important legacy of high $T_c$ cuprates is that the mean field theory like the Landau theory of Fermi 
liquid\cite{landau1,landau2,landau3} 
and the BCS theory of superconductivity\cite{BCS} works in the quasi-two dimensional system with strong electron 
correlations\cite{path,hussey}.
Of course the Fermi liquid theory as formulated by Landau for a spherical Fermi surface cannot be applied  directly to high $T_c$ 
cuprates. In particular the quasi-two dimensionality and the resulting nesting feature of the Fermi surface has to be considered. But 
this feature is readily handled in terms of the renormalization theory of two dimensional Fermi liquid\cite{shankar,metzner,houghton}. 
Also as to the 
superconductivity the one band Hubbard model will give the simplest starting point. Then as discussed by Scalapino and others, 
d$_{x^2-y^2}$ superconductivity follows immediately\cite{scalapino,pao}. Further if one limit oneself to  single crystals of optimally 
doped high $T_c$ 
cuprates, one can do quantitative test of the BCS theory of d-wave superconductor in the weak coupling 
limit\cite{d-wave,hussey,universal}. Unfortunately until now 
only three kinds of single crystals of high $T_c$ cuprates are available: LSCO, YBCO and Bi2212. If you compare 
$\Delta(0)/k_BT_c=2.14$, 
the weak coupling theory prediction\cite{d-wave} for d-wave superconductor ($\Delta(0)$ is the maximum of the energy gap) to the one 
obtained for 
the optimally doped single crystals, we obtain 2.14, 2.8, 5 for LSCO, YBCO and Bi2212 respectively. This means that the 
superconductivity in LSCO is very close to the weak coupling limit, the one in YBCO is moderately in the strong coupling limit, while 
Bi2212 is definitely in the strong coupling limit\cite{szummad-wave}. In Fig. \ref{phase} a generic phase diagram of the hole doped 
high $T_c$ cuprates is 
shown. It is still controversial where $T^*$ line hits the superconducting transition temperature curve $T_c$. But from the validity of 
the mean field theory at optimal doping we assume that it hits somewhat in the underdoped side. Then it is possible that the extension 
of this line continues to $T=0$~K at $x=0.15$ at the quantum critical point. On this point we may refer to an earlier resistivity 
measurement in high magnetic field though it is limited unfortunately to only LSCO system\cite{ando}.
Therefore the d-wave superconductivity in high $T_c$ cuprates is well understood in terms of two dimensional one band Hubbard model 
except one caveat: what means $T^*$? Earlier it was believed that $T^*$ is a crossover temperature where either superconducting or 
antiferromagnetic fluctuations becomes important\cite{ybco2}. More recently possible phase transition to d-wave density wave at 
$T=T^*$ has been 
proposed\cite{benfatto,nayak,app}. The most serious objection to this model is that no jump in the specific heat at $T=T^*$ has been 
observed until now, though 
many physical quantities like nuclear spin lattice relaxation rate $T_1^{-1}$, magnetic susceptibility, electric conductivity exhibit 
kinks at $T^*$\cite{tallon}.
The d-wave nature of density wave has been established by angular dependent photo-electron spectrum study\cite{timusk}. Another less 
indirect 
signature of d-wave is the surprising relation $\Delta(0)/T^*=2.14$ (the weak coupling result for d-wave density wave as well as 
for d-wave 
superconductor\cite{nagycikk,d-wave}) established by STM study of $\Delta(0)$ (the energy gap in the density of states at $T=0$~K) in 
LSCO, YBCO and Bi2212\cite{renner}.
Therefore the only remaining question is if it is UCDW or USDW. We have proposed recently that USDW can interpret very readily two 
crucial experiments observed in the pseudogap phase in high $T_c$ cuprates YBCO and Bi2212: the weak antiferromagnetism\cite{sidis} and 
the optical 
dishroism in ARPES\cite{kaminski}.
Sidis et al. observed the appearance of the weak antiferromagnetism at $T=T^*$. This feature is qualitatively very similar to the weak 
antiferromagnetism observed in URu$_2$Si$_2$\cite{roma}. Unfortunately the temperature dependence of the intensity of the AF amplitude 
is rather 
different from the one in URu$_2$Si$_2$. But there are a few more possible contributions what we have neglected.
In this picture the spin configuration of USDW is given by $S^\pm=S_x\pm iS_y$ lying in the $a-b$ plane. There are many attempts to 
describe this feature in terms of orbital angular momentum, but these models look too artificial. Perhaps the optical dishroism 
observed in the pseudogap phase in Bi2212 is still more controversial\cite{kaminski}. Indeed this is predicted by Chandra Varma based 
on a three band 
Hubbard model with a complicated order parameter\cite{varma}. There are many works trying to reinterpret this feature in terms of 
orbital currents 
associated with d-wave density wave\cite{nguyen}.

One of the natural consequence of d-wave SDW with spin component $S^\pm$ is the optical dichroism as observed by Kaminski et 
al.\cite{kaminski}.
The fact that the spin component lies in the $a-b$ plane is consistent with neutron scattering experiment\cite{sidis}. Making use of 
the standard 
procedure to calculate ARPES, we find
\begin{equation}
I_\pm\sim1\pm\frac{\Delta({\bf k})}{E({\bf k})}\label{nyolc}
\end{equation}
or
\begin{equation}
P=\frac{I_+-I_-}{I_++I_-}=\frac{\Delta({\bf k})}{E({\bf k})},
\end{equation}
where $\Delta({\bf k})=\Delta\cos(2\phi)$ and $E(\bf k)$ is the QP energy. Eq. (\ref{nyolc}) tells  that the optical dichroism is 
proportional to $\cos(2\phi)$.
In particular $P=0$ for $\bf k$ in the nodal directions while $P$ takes the maximum value at the antinodal directions.
These facts are consistent with experiment\cite{kaminski}. We expect also in a uniform ground state the 100 \% dichroism.
But small dichroism is mostly due to the nonuniform ground state. We further expect the spin polarization of the outcoming electrons 
parallel to the photon polarization.

Also we propose that the angular dependent magnetoresistance will be a powerful method to investigate the d-wave density wave in high 
$T_c$ cuprates\cite{makifra}. In a magnetic field $\bf H$ applied as shown in Fig. \ref{fermisurf} (after replacing $a\rightarrow b$, 
$b\rightarrow c$, $c\rightarrow a$), the QP spectrum in d-wave density 
wave 
changes to
\begin{equation}
E_\pm=\sqrt{2\sqrt{2}e|H|\Delta(v_Fa|\cos(\theta)|+v_cc\sin(\theta)|\sin(\phi\pm\frac\pi
4)|)},
\end{equation}
where $v_F$ and $v_c$ are the Fermi (in-plane) and perpendicular velocity,
respectively.
Again we followed Ref. \cite{Ners1} and neglected the imperfect nesting terms for simplicity. Therefore the magnetoresistance is given 
by
\begin{equation}
\frac{\rho(H,\theta,\phi)-\rho(0,\theta,\phi)}{\rho(0,\theta,\phi)}=\frac{e^{\beta(E_++E_-)}-1}{e^{\beta
E_+}+e^{\beta E_-}+2}
\end{equation}
A typical $\theta$ dependence is shown in Fig \ref{phase}.

Although this cannot distinguish between d-wave CDW and SDW, at least this will provide a unique test of the UDW proposed in high $T_c$ 
cuprates.

Very recently d-wave symmetry of the superconductivity in heavy fermion layered compound CeCoIn$_5$ and in the organic superconductor
$\kappa$-(BEDT-TTF)$_2$Cu(NCS)$_2$ have been established\cite{izawa3,izawa4}. Further both of these superconductors lie in the 
vicinity of a kind of 
antiferromagnetic state (most likely a kind of SDW)\cite{pinteric,muller}. Perhaps the most surprising phenomenon is the dependence of 
superconductivity in 
$\kappa$-(BEDT-TTF)$_2$Cu[N(CN)$_2$]Br on the cooling rate\cite{pinteric}. For clarity we consider two extreme cases: the well annealed 
crystals are 
kept at liquid N$_2$ temperature for three days before final slow cooling to the temperature region around 10~K, while the quenched 
crystals were cooled down to the liquid He temperature within one hour. Surprisingly the superconducting transition temperature is 
little affected by the different cooling procedure.
But from the diamagnetic response it is shown that the superfluid density in the quenched sample is less than 1\% of the annealed 
sample. Further the temperature dependence of the superfluid density of the annealed sample is consistent with the one in d-wave 
superconductor, while the one 
for the quenched sample can be interpreted in terms of the one in s-wave superconductor. Therefore we suspect that the origin of the 
controversy over d-wave versus s-wave superconductivity lies in the question of the cooling rate. It is well understood that disorder 
in the ethylene groups attached to the BEDT-TTF molecule is destructive to superconductivity, though we do not know how. 
The slow cooling through the glassy transition temperature (100~K-70~K) where the ethylene group disorder sets in, helps to form more 
ordered ethylene groups\cite{muller}. Also it is very likely that disorder in the ethylene group is more disastrous to 
superconductivity than to 
SDW. Then a natural question is if this kind of SDW is USDW or not. Unfortunately, there is no experimental data on the 
characterization of this antiferromagnetic order parameter.
Therefore we are sure that ADMR will be very useful to clarify this question.

Also can the weak superconductivity or gossamer superconductivity\cite{laughlin} found in the quenched sample be described in terms of 
coexisting d-wave superconductivity and d-wave SDW?
We believe this is one of the most interesting questions in organic superconductors.

\section{Concluding remarks}

We have seen that UCDW and USDW are very likely realized in organic conductors, in heavy fermion systems and in the pseudogap phase in 
high $T_c$ cuprates. Also we have proposed that the angular dependent magnetoresistance will provide a unique probe to discover UDW. In 
particular we have identified successfully UCDW in $\alpha$-(BEDT-TTF)$_2$KHg(SCN)$_4$. Also we have pointed out that there are many 
similarities among the pseudogap phase in high $T_c$ cuprates, the glassy phase in organic superconductor $\kappa$-(BEDT-TTF)$_2$X and 
the 115 compounds in heavy fermion systems including CeCoIn$_5$ and PuCoGa$_5$\cite{sarrao}.
The latter system with superconducting transition temperature $T_c=18$~K is of great interest. Also as unconventional 
superconductivity becomes the superconductivity of the 21st century, we are confident that UCDW and USDW will be the density wave in 
this new century.

\begin{theacknowledgments}

We are very much pleased to dedicate this work for the 60th birthday
of Professor Mancini, our friend and our colleague. We wish for Nando a number of coming fruitful years. Also we thank
Fernando for founding the training school in true Platonic
tradition, which provides us a small civilized corner in the present
turbulent universe.
We thank Mario Basleti\'c, Bojana Korin-Hamzi\'c, Amir Hamzi\'c, M. V. Kartsovnik, Marko Pinteri\'c, Silvia Tomi\'c and Peter 
Thalmeier for discussions and collaborations on related subjects.
This work
was supported by the Hungarian National Research Fund under grant numbers
OTKA T032162 and  TS040878.
\end{theacknowledgments}

\bibliographystyle{aipprocl} 
\bibliography{makisaler.bib}

\end{document}